# Flexible Transparent High-Voltage Diodes for Energy Management in Wearable Electronics

Yonghui Zhang[a,b], Zengxia Mei[a,*], Tao Wang[a], Wenxing Huo[a,b], Shujuan Cui[a,b], Huili Liang[a], and Xiaolong Du[a,b,*]

[a] Beijing National Laboratory for Condensed Matter Physics, Institute of Physics, Chinese Academy of Sciences, Beijing 100190, P. R. China
[b] School of Physical Sciences, University of Chinese Academy of Sciences, Beijing 100049, P. R. China

Corresponding authors at: Beijing National Laboratory for Condensed Matter Physics, Institute of Physics, Chinese Academy of Sciences, Beijing 100190, P. R. China
E-mail addresses: zxmei@iphy.ac.cn (Z. Mei) and xldu@iphy.ac.cn (X. Du)

This work reports flexible fully transparent high-voltage diodes that feature high rectification ratio ($R_r \sim 10^8$) and high breakdown voltage ($V_b \sim 150$ V) simultaneously, combined with their applications as building blocks of energy management systems in wearable electronics where triboelectric nanogenerators (TENGs) are used as power source. Both experimental results and technology computer aided design (TCAD) simulations suggest that $R_r$ and $V_b$ can be modulated by the offset length in an opposite tendency. The low reverse leakage current ($\sim 10^{-15}$ A/μm) guarantees an ultra-low power consumption in standby mode, which is a core issue in wearable device applications. Besides the unprecedented electrical performance, the diodes exhibit good mechanical robustness with minimal degradation throughout the strain and fatigue tests. By incorporating these high-voltage diodes into half-wave and full-wave rectifier circuits, the high alternating current (AC) output voltage of TENGs is successfully rectified into direct current (DC) voltage and charged into supercapacitors (SCs), indicating their high integration and compatibility with TENGs, and thus their promising applications in various wearable electronic systems.

## 1. Introduction

Wearable electronics[1,2] has advanced at a rapid pace in the past dozen years, driving a fascinating transformation of the consumer electronics pattern. The well-functioning wearable devices strongly depend on a stable supply of power energy. The self-powered wearable electronics which possesses good coordination among energy harvesting, management, and storage is highly desired because it is light weighted, highly integrated, and free of external electrical connections. Triboelectric nanogenerator (TENG), a kind of light and compact device which harvests electrical energy from ambient motions, is regarded as the perfect power source in wearable electronics[3–9] ever since the introduction of flexible TENG in January 2012[10]. Up to now, high area power density of 500 W/m$^2$[11] and high instantaneous conversion efficiency of 70.6%[12] have been achieved, which are sufficient for the power demands of many portable gadgets. As for the flexible energy storage devices[13], great progresses have been made in both flexible lithium-ion batteries (LIBs)[14,15] and supercapacitors (SCs)[16,17]. The present difficulty of applying TENGs in wearable electronics lies in its energy management for their large impedance and unbalanced load matching. Effective power management plays a crucial role in practical energy utilization, which has always been the major headache and bottleneck for the practicability of TENG[3,5,7,18,19]. To transform the high alternating current (AC) output voltage of TENGs into usable low direct current (DC) voltage, flexible high-voltage rectifiers and converters are indispensable units wherein flexible high-voltage diodes are one kind of core components[20].

Wide band-gap semiconductors own superiority to silicon (Si), germanium (Ge), and organic materials in high-voltage devices due to their much higher breakdown field strength and better thermal stability[21,22]. Although high-voltage diodes based on silicon carbide (SiC)[23], gallium nitride (GaN)[24], and diamond[25] have made progresses both in academic and industrial areas, they are generally fabricated at elevated temperature conditions, which is not compatible with the flexible electronic techniques on polymer substrates. On the contrary, wide band-gap oxide semiconductors, more specifically, zinc oxide (ZnO)-based materials, have been extensively used in flexible electronics due to their good electrical performance and low processing temperature[26]. ZnO is a "rediscovered" semiconductor receiving remarkable interest on behalf of its unique merits and promising technological applications. Currently, the flexible diodes are all fabricated below 200 °C, using either low-temperature-synthesized oxide[27–35] and organic[36–51] materials or high-temperature-prepared Si[52–63] and Ge[64] materials combined with transfer method[65], as summarized in **Figure 1**. The highest reverse voltage or breakdown voltage ($V_b$) of these flexible diodes are no more than 20 V, with an exception in References 55 and 59 where flexible single-crystalline Si wafers (30 μm thick) were used as the active layers. Deficiently, only 0.13% strain can be withstood due to such a large thickness. The other devices with thinner active layers (~100 nm thick) are not expected to sustain such high voltage because high electric field (~MV/cm) could easily breakdown the junctions or bulk materials, and that's why no real flexible high-voltage diode has been reported yet.

In this paper, we report flexible transparent high-voltage diodes with $V_b$ as high as 150 V, rectification ratio ($R_r$) over $10^8$ and reverse leakage current ($I_r$) as low as $10^{-15}$ A/μm, plus the demonstration of half-wave and full-wave rectifier circuits based on the high-voltage diodes components. The device working mechanism was studied via current-voltage (*I-V*) characteristics and technology computer aided design (TCAD) simulations. Both of the electrical and mechanical tests exhibit their practical usability in flexible electronics. Lastly, half-wave and full-wave rectifier circuits, built out of a single diode and four-diode bridge, respectively, were used to rectify high AC voltage and charge SCs, manifesting its potential serviceability in future wearable electronic equipments.

**2. Results and Discussion**

*2.1 Device fabrication and electrical characterization*

The commercialized manufacture of wearable electronics requires low-cost and fast-speed microfabrication technologies. We utilize the conventional lithography technique and film synthesis methods which suit for mass production of flexible devices. The fabrication procedure and details can be seen in **Figure 2**a and Experimental Section, respectively. The cross-sectional view (Figure 2b) shows the structure diagram of the high-voltage diodes. Cathode and top-anode (T-anode) electrodes are located on the edge and at the center of the circle, respectively, forming ohmic contacts to ZnO channel. The bottom-anode (B-anode) is set at the bottom of the center, forming a metal-insulator-semiconductor (MIS) capacitor with ZnO and alumina ($Al_2O_3$) to control the conductivity of the channel area above it. Compared with the conventional field-effect diode (FED) architecture[27], an offset region is introduced between B-anode and cathode to withstand higher voltage as it does in high-voltage transistors[66–69]. The photograph of a completed 25×25 mm plastic wafer and the scanning electron microscope (SEM) image of a high-voltage diode are shown in Figure 2c and 2d, respectively.

The reverse breakdown voltage is one of the most important parameters of the high-voltage diodes which signifies the safe operation range of the resulting devices. **Figure 3**a presents the *I-V* characteristics of a flexible high-voltage diode with a 5 μm-offset region, which shows a high $V_b$ of 147 V. The statistical data of $V_b$ obtained from 33 devices is shown in the histogram (Figure 3b), with an average $V_b$ of 131.21 V and standard deviation ($\sigma$) of 10.83 V. $R_r$ is another important parameter which depicts the ability of unidirectional conductivity in diodes. A high $R_r$ of $4.4 \times 10^8$ was obtained at ± 40 V in our flexible high-voltage diode (Figure 3c), 3~4 orders of magnitude higher than that of most flexible diodes. The average of log ($R_r$) and standard deviation obtained from 41 devices were 6.70 and 0.64, respectively (Figure 3d). Furthermore, an $I_r$ as low as $10^{-15}$ A/μm promises the ultra-low power consumption, which is desirable for wearable device applications[70]. The astonishing performance of the presented flexible high-voltage diode guarantees its potential applications in high-efficiency energy management systems.

To explore the dependence of $V_b$ and $R_r$ on the offset length, two kinds of diodes with offset length of 10 μm and overlap length of 2 μm were also fabricated in the same batch for contrast. The 10 μm-offset diode does not breakdown under a reverse bias of 210 V (the maximum output voltage of Keithley 4200 semiconductor characterization system), but, as a trade off, the $R_r$ reduces to $3.38 \times 10^3$ (Figure 3e). On the contrary, the 2 μm-overlap diode shows a higher $R_r$ and lower $V_b$. Intriguingly, there is an opposite trend of $V_b$ and $R_r$ with regard to the offset length. For diodes with a longer offset region, higher $V_b$ and smaller $R_r$ could be obtained; and vise verse (Figure 3f). This phenomenon is analogous to that in flexible ZnO-based high-voltage transistors [66,68,69].

Note that the presented electrical performance in Figure 3 is based on the devices with a nominal offset value of 5 μm. However, due to the unavoidable misalignment issue during the lithography process, the minimum offset length might be reduced to around 3 μm. Besides, this value might vary from device to device, leading to an inhomogeneity that cannot be ignored, as described in Figure 3b and 3d. Currently, we are working on a self-aligned lithography technique aiming for the solution to the misalignment problem and hence the improvement of device uniformity and repeatability.

The maximum rectifying frequency is the most important parameter which limits the use of devices in high-frequency applications. Restricted by the significantly high equivalent resistance (~GΩ) and the maximum output voltage of Keithley 3390 Arbitrary Waveform/Function Generator we employed (10 V), the voltage dropped on the oscilloscope probe (10 MΩ) is too small to be detected. As an alternative, transient response measurements of the presented high-voltage diodes were conducted. The high-voltage square wave was successfully rectified into DC voltage as shown in Figure 3g (Inset shows the rectification circuit diagram.). The rising and falling times (between the minimum and maximum) are about 400 μs and 2.2 ms, respectively. After comparing these data with the results of the conventional FED[27] (Figure 3h and 3i), the rising and falling edges of the HV diode were found slightly sharper. Therefore, the maximum operation frequency should be higher than that of 1 MHz in conventional FEDs. The killer application of TENG, which the presented high-voltage diodes are designed to work with, is in harvesting low-frequency energy such as ocean wave energy and body movements energy. The typical frequency is less than 5 Hz[9,19]. In this case, the presented high-voltage diodes are competent candidates in combination with TENGs in weareable electronics.

*2.2. TCAD simulation of device working principles*

The working principle of the unidirectional conductivity in FED was depicted in detail in our previous report[27]. Conductive channel is formed by accumulating electrons near the ZnO/Al$_2$O$_3$ interface under a positive bias, leading to a large positive current, while the conductive channel is shut down by depleting electrons from the ZnO/Al$_2$O$_3$ interface under a negative bias, causing a negligible reverse current. In this report, the effect of an offset region on the electrical performance of FED was specifically introduced and emphasized, plus the influence of offset and overlap length on high-voltage operation features. The simulation was carried out by atlas simulator included in the Silvaco TCAD software, to derive the electrical parameters under positive and negative biases in various device architectures, respectively. Two kinds of FED devices were defined (**Figure 4**a): (i) conventional FED with an overlap length of 1 μm, 2 μm and 3 μm, respectively, and (ii) high-voltage FED with an offset length of 1 μm, 2 μm and 3 μm, respectively. Simulated *I–V* characteristics of the above-mentioned devices (Figure 4b) show an analogous tendency to the experimental results in Figure 3e. The increase of offset length results in less control of voltage over current in the channel, i.e., the larger the offset length is, the lower the forward current is and the higher the reverse current is. Figure 4c shows the electron concentration (*n*) profiles of ZnO layers in these six structures along cutline 1 (height = -0.121 μm, lateral = 0-8 μm) under a +40 V voltage bias. The ZnO channel layers can be divided into two regions with high electron concentration and low electron concentration, respectively, corresponding to the gated and ungated or offset regions in Figure 4d[67]. Increasing the offset length actually means decreasing the gated length, assuming a constant overall channel length. While the gated region equals to a conventional FED whose resistance is controlled by anode bias, the ungated or offset region can be equivalent to a resistor whose resistance is independent of the applied bias. Under a positive bias, electrons are accumulated in the gated region, casuing an increase of its conductivity, while the electron concentration and resistivity of the offset region remains unchanged. In this case, the conductance of the ungated or offset region as well as the bias dropped on the gated region decreases with longer offset length, leading to the lower forward current, as shown in Figure 3e and Figure 4b. On the other hand, when the device is negatively biased, electrons are depleted in the gated region, making it highly resistant, remarkably different from the still unchanged electron concentration and resistivity of the ungated or offset region. Therefore, the reverse current will be higher with a longer offset length, as seen in Figure 3e and Figure 4b.

The offset region is indispensable in high-voltage diodes, although it results in lower forward current and higher reverse current. In overlap structures, the high reverse voltage of 150 V is distributed vertically within the 150 nm length (50 nm-thick ZnO on 100 nm-thick $Al_2O_3$) region between B-anode and cathode, which generates a high electric field strength ($E$) of more than 10 MV/cm in $Al_2O_3$ layer as can be seen in Figure 4e (extracted along cutline 2: height=-0.021 μm, lateral=0-8 μm). Such a high electric field can easily cause a breakdown both at interfaces and in bulk materials. The hazardous electric field will be mitigated by introducing an offset region between B-anode and cathode electrodes. There are spikes in the electric field profiles, locating near the electrode edge in all of these offset structures (Figure 4f). The spikes are the weak points (with maximum electric field values) in the devices, from where the breakdown first occurs. In our next work, we will focus on the diminution or elimination of the spikes to achieve a desirable higher-voltage operation performance.

*2.3. Mechanical property evaluation*

Flexible devices are designed to maintain their electrical performance throughout intensive and multiple mechanical bending actions. To evaluate the feasibility of present high-voltage diodes in wearable electronics, *I-V* characteristics is conducted while the substrate is flat and bent at various radii (*r*), respectively (**Figure 5**a). Minimal degradation in device performance was observed when the substrate was bent till to $r = 11$ mm, corresponding to a tensile strain ($\varepsilon$) of 0.57%. However, the forward current was dramatically reduced to $\sim 10^{-14}$ A when $r = 10$ mm ($\varepsilon = 0.63\%$) due to the crack of the electrode contact pads as observed through a microscopy (not shown here). After relaxing the substrate to the flat state, the crack was converged together and the forward current recovered the previous level. What is more, the reverse current decreased a little bit under flexure and after relaxation, leading to a relatively higher $R_r$ compared to the one in flat state ($R_{r0}$) (Figure 5b). The fatigue property of the flexible high-voltage diodes was further tested with a self-assembled slide table system as shown in the inset photograph of Figure 5d. The electrical performance did not show noticeable degradation after enduring 100 times tensile strain of 0.52%. After 10000 times bending, the reverse current became unstable at a voltage around 100 V (Figure 5c). Even more impressive, $R_r/R_{r0}$ keeps almost unchanged after these bending actions (Figure 5d). The mechanical robustness of the flexible high-voltage diodes manifests its potential to be applied in the wearable electronics and it could be improved by further optimizations, such as neutral surface encapsulation[71] and buffer layer insertion[72].

*2.4 Half-wave and full-wave rectifier circuits*

*2.4.1. Half-wave rectifier circuits*

Half-wave rectifier circuit is constructed with one single flexible high-voltage diode component to demonstrate its practical usability of converting AC voltage to DC voltage (**Figure 6**a). The input AC voltage ($V_{IN}$) was acquired from the electric grid (220 V, 50 Hz, national standard voltage in China) through an isolation transformer and a step-down transformer. The output DC voltage ($V_{OUT}$) was recorded with an oscilloscope through passive voltage probes, whose input resistance and capacitance were used as the load resistance ($R_L$) and probe load capacitance ($C_{Lp}$) in the rectifier circuits, respectively. The sinusoidal $V_{IN}$ with an amplitude ranging from 40 V to 110 V was applied at the input terminal, while the half-wave $V_{OUT}$ was obtained at the output terminal with $1 \times$ and $10 \times$ attenuation configurations (Figure 6b). (See Experimental section for details on the probe configuration.) The $V_{OUT}$ amplitude can be modulated through the variation of $V_{IN}$ and $R_L$. As seen in Figure 6c, both the $V_{OUT}$ values with 10 MΩ and 1 MΩ load increase monotonously with $V_{IN}$ and the former $V_{OUT}$ is higher than the latter. Finally, the $V_{OUT}$ with 1 MΩ and 10 MΩ load was successfully charged into 1 μF and 10 μF SCs with sinusoidal $V_{IN}$ of 50 V amplitude, respectively. The charge and discharge processes follow the resistance-capacitance circuit rules (time constant $\tau = \sqrt{R_L C_L}$, where $C_L$ is the total capacitance of the probe and the SC), which means the circuit with a smaller $R_L$ and $C_L$ will be charged and discharged faster than that with a larger $R_L$ and $C_L$ (Figure 6d). The capacitor voltage is saturated at 2.64 V and 7.76 V with 1 MΩ and 10 MΩ load in 1 μF SC, respectively, which are much smaller than the $V_{OUT}$ amplitude of 8.00 V and 19.80 V at $V_{IN}$ = 50 V in Figure 6c. That is because the charge process only occurs when $V_{OUT}$ is higher than the SC voltage. In another word, the SC is actually discharged during the time period in which $V_{OUT}$ is lower than the SC voltage (see the inset of Figure 6d).

*2.4.2. Full-wave rectifier circuits*

The TENGs generate high AC voltage, which consists of positive and negative parts in one cycle. To make full use of the electrical power, full-wave rectifier is achieved with diode bridge constituted by four high-voltage diodes which were fabricated by the same process as the one for single high-voltage diode shown in Figure 2a and connected accordingly using the built-in micro-wires (**Figure 7**a). The full-wave rectifier circuit is shown in Figure 7b, with various AC square wave voltages as $V_{IN}$ (Figure 7c). After connected to the circuit with 10 MΩ load, the AC $V_{IN}$ is successfully transformed into square DC voltage (Figure 7d). However, the $V_{OUT}$ levels obtained from the positive and negative parts of $V_{IN}$ are not necessarily the same (inset of figure 7d) due to the inhomogeneous performance of the four diodes (Figure S1, Supporting Information). Figure 7e illustrates that both the high and low $V_{OUT}$ levels increase monotonously with $V_{IN}$. A square AC voltage of 50 V is utilized as $V_{IN}$ to charge 1 μF and 10 μF SCs with 1 MΩ load (Figure 7f). Unlike the case in the half-wave rectifier circuit (Figure 6d), the saturation voltage exhibits a more pronounced jagged characteristic in 1 μF SC. The reason is that the SC is charged with high $V_{OUT}$, and followed by a subsequent discharge with low $V_{OUT}$, as shown in the inset of Figure 7f. This phenomenon turns to be less prominent in 10 μF SC because of the relatively large $\tau$ (Figure 7f), and becomes completely unnoticeable at higher-frequency $V_{IN}$ (Figure 6d). To verify the generality and practicability of the rectification circuits, triangle-wave high AC voltage is also used as $V_{IN}$, the experimental results can be found in Figure S2 (Supporting Information). A screen capture video of a typical test procedure with a self-compiled Labview program is shown in Video S3 (Supporting Information). Besides, full-wave rectification is achieved with the triangle-wave high AC voltage while the substrate was both in flat and bent status (Figure 7g), demonstrating its practical serviceability in wearable systems.

*2.5 Fully transparent high-voltage diodes*

Investigation on transparent and especially fully transparent electronics is of great interests for diverse applications, such as see-through displayers and transparent sensors for security applications. As for wearable electronics, the aesthetic property of the fully transparent circuits makes the wearable products pretty, simple and in fashion. Flexible fully-transparent high-voltage diodes are fabricated by utilizing indium tin oxide (ITO) as the transparent electrodes (inset of **Figure 8**a). The completed device, including the substrate, is optically transparent in visible spectrum range with a high transmittance over 80% (Figure 8a). The *I-V* characteristics shows high $R_r$ of $2.18 \times 10^8$ and high $V_b$ of 145 V, comparable to the above-mentioned diode with opaque chromium (Cr) electrode (Figure 8b). To evaluate its practical usability in combination with high-voltage power source, random high AC voltage, generated from electromagnetic generator (EMG) (Figure 8c) and TENG (Figure 8e), is involved as $V_{IN}$. The random high AC voltage is successfully rectified into DC voltage through half-wave (Figure 8d) and full-wave (Figure 8f) rectifier circuits, demonstrating the compatibility of the presented high-voltage diodes with EMGs and TENGs. Video S4 (Supporting Information) shows the processes of the measurements of $V_{IN}$ and $V_{OUT}$ of full-wave rectifier circuits with TENG as power source. As a complete energy supply strategy, the charge of capacitors with diode bridge through rectifying electrical power form TENG is achieved (Figuer 8g). The capacitors are charged by the diode bridge at the first stage. After a certain amount of palm tappings, the voltage of the capacitors is measured by an oscilloscope. The charging and testing processes are conducted separately, because the capacitors are easily discharged during the measurements. After 500 times manual palm tappings, the voltage reaches 5.6 V, 1.48 V and 0.2 V for 1 nF, 10 nF and 1 µF capacitors, respectively (Figure 8h). This complete power supply strategy demonstrates the serviceability of the presented flexible transparent high-voltage diode in combination with TENGs in wearable electronics.

3. **Conclusion**

In conclusion, we have presented flexible transparent high-voltage diodes as building blocks for energy management systems in wearable electronics. High $R_r$ of $4.4 \times 10^8$ and high $V_b$ of 147 V are achieved simultaneouly, far superior to previous flexible diode reports. Experimental and TCAD simulation results reveal that $R_r$ and $V_b$ closely relate to the offset length in an opposite trend. The devices show good robustness in mechanical strain and fatigue tests under flexure. High AC voltage is rectified into DC voltage, through half-wave and full-wave rectifier circuits composed with our high-voltage diodes, and successfully charged into SCs. Furthermore, our high-voltage diodes demonstrate well compatibility and integration possibility with TENGs. Based on the above evidences, the flexible transparent high-voltage diodes are believed to be of great potential in energy management systems of the rapidly developed wearable electronics.

## 4. Experimental Section

*4.1 Device fabrication*

The fabrication procedure is illustrated in Figure 2a. Step 1: polyethylene naphthalate (PEN) substrate (125 μm thick) was ultrasonically cleaned in acetone and isopropyl alcohol and blown dry with nitrogen. After then, the substrate was put into atomic layer deposition (ALD) chamber and degassed at 110 °C for 12 hours. Subsequently, $Al_2O_3$ (50 nm thick) was deposited to encapsulate the substrate without breaking the vacuum. Step 2: a bottom chromium (Cr, 50 nm thick) or indium tin oxide (ITO, 120 nm thick) anode was deposited by radio frequency (RF) magnetron sputtering and patterned by UV-lithography via lift-off. Step 3: an $Al_2O_3$ dielectric layer (150 nm thick) was deposited by ALD at 100 °C and patterned by UV-lithography followed by wet etching in phosphoric acid ($H_3PO_4$ 80%, for 50 s at 70 °C). Step 4: a ZnO active layer (50 nm thick) was deposited by RF-sputtering with a ceramic target (99.99%) at room temperature and patterned by UV-lithography followed by wet etching in hydrochloric acid (HCl 1%, for 3 s at 25 °C). Step 5: the same ITO sputtering and lift-off process was used to define the top contacts. The full-wave rectifiers were fabricated in the same process along with the single diode devices using different patterns on the same mask.

*4.2 Device characterization*

*I-V* characteristics measurements in Figure 3, Figure 5 and Figure 8 were performed in dark air at room temperature using source-measurement unit included in Keithley 4200 semiconductor characterization system. The transient response measurements in Figure 3g-3i were performed with high-voltage square wave (with an amplitude of 110 V) generated from Keithley 2400 source meter as power source and 10 MΩ resistor as load. The voltage was recorded by a Tektronix TDS1000B-SC oscilloscope via P2220 passive voltage probes. The mechanical bending test in Figure 5b was conducted using a clamp with precise screw scale. The strain value was calculated through $\varepsilon = d/(2r)$, where $d$ is the thickness of the substrate and $r$ the radius of the curvature. $r$ was calculated using geometric relations of circular arcs: $L = 2\theta r$ and $D = 2r\sin\theta$, where $L$ is the arc length, $D$ the chord length, and $\theta$ half of the arc angle[26]. The fatigue test in Figure 5d was conducted with a self-assembled slide table system composed of a stepping motor, a motor diver, a digital controller, a power source, a limiting stopper and a slide table. One bending cycle takes approximately 1.5 seconds. In Figure 6, the $V_{IN}$ was taken from national electrical grid through an isolation and a step-down transformer and $V_{OUT}$ was measured by a Tektronix TDS1000B-SC oscilloscope via two P2220 passive voltage probes ($R_L$ = 1 MΩ, $C_{Lp}$ = 110 pF at 1 × attenuation position; $R_L$ = 10 MΩ, $C_{Lp}$ = 17 pF at 10 × attenuation position). In Figure 7 and Figure S2, the $V_{IN}$ was taken from a Keithley 6487 picoammeter. The transmittance spectra in Figure 8 was obtained with a SHIMADZU UV 3600 plus spectrophotometer. Radom high AC voltage in Figure 8c was obtained by connecting radom AC voltage, generated from electromagnetic induction effect, to a step-up transformer.

*4.3 TCAD simulation*

The simulations in Figure 4 were performed by ATLAS simulator included in the Silvaco TCAD software. A nonlinear mesh was defined to accurately characterize the parameters in active areas while coarsely elsewhere. On this basis, a ZnO layer (50 nm thick, energy gap $E_g$ = 3.37 eV, affinity $\chi$ = 4.5 eV) and an $Al_2O_3$ layer (100 nm thick, dielectric constant $\varepsilon_r$ = 9.3) were defined to serve as channel and dielectric layer, respectively, with conductor (work function $\Phi_M$ = 4.33 eV) serving as contact electrodes. Fermi–Dirac statistics model was used to get a precise description of electrons in thermal equilibrium.

4.4 *Fabrication of TENGs*

The TENGs in Figure 8 and Video S4 (Supporting Information) operate through the vertical contact-separation mode with polyethylene terephthalate (PET)/ITO foil coated with polydimethylsiloxane (PDMS) as the upper contact material and aluminum foil as the lower contact material.

**Appendix A. Supporting information**
Supplementary data associated with this article can be found online or from the author.

**Notes**
The authors declare no competing financial interest.

**Acknowledgements**
This work was supported by the National Natural Science Foundation of China (Grants No. 11674405, 11675280, 11274366, 51272280, and 61306011).

The authors thank Prof. Yicheng Lu (Rutgers University, USA) for inspirational suggestion and discussion, and acknowledge support from the Laboratory of Microfabrication in Institute of Physics, Chinese Academy of Sciences.

**Figures**

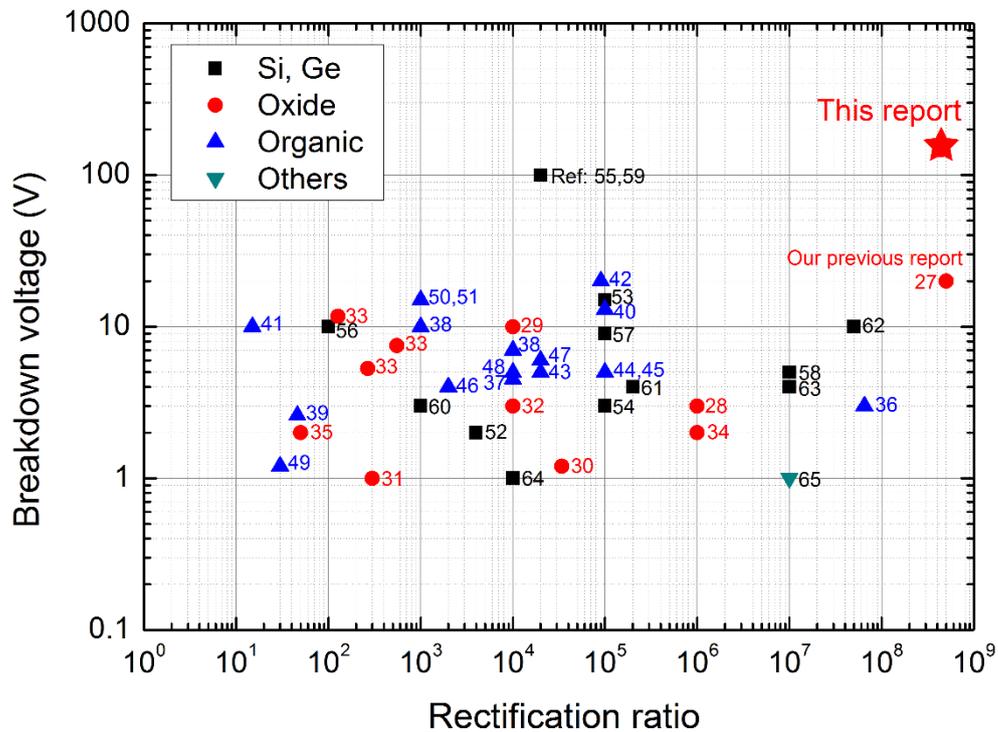

**Figure 1.** Summary of the reports on flexible diodes from year 2007 till now. Field-effect diodes in our previous work[27] and high-voltage diodes in this work show better performance both in $V_b$ and $R_r$.

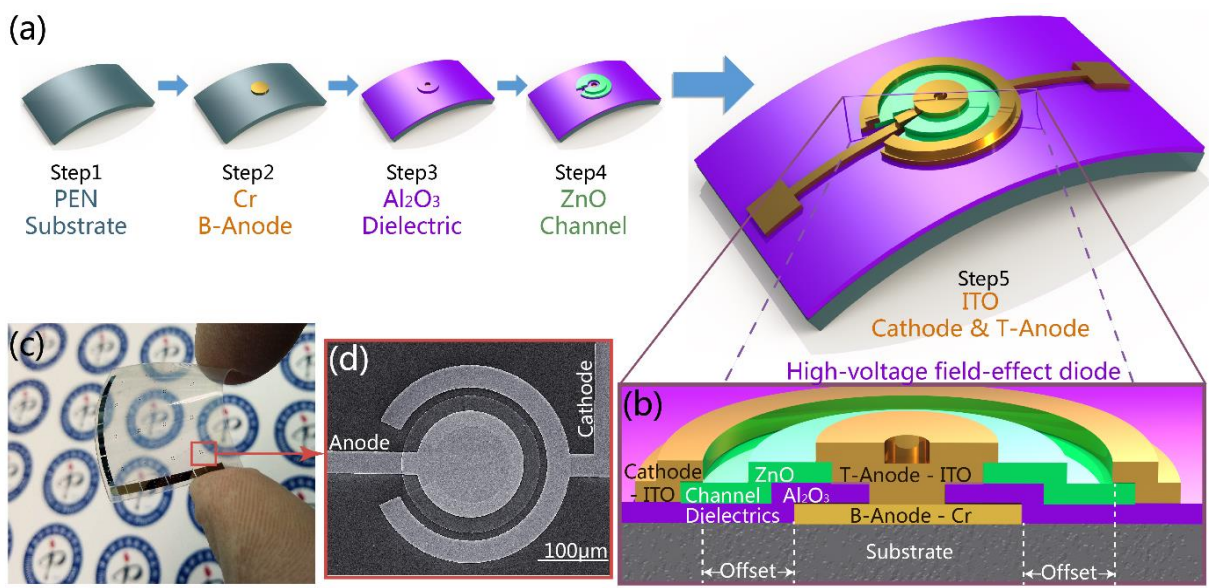

**Figure 2.** Fabrication procedure, device structure, and SEM image of the high-voltage diodes. (a) Process flow of a typical fabrication procedure. (b) Cross-sectional view of a high-voltage diode. (c) Photograph of a completed 25×25 mm plastic wafer. (d) SEM image of a high-voltage diode.

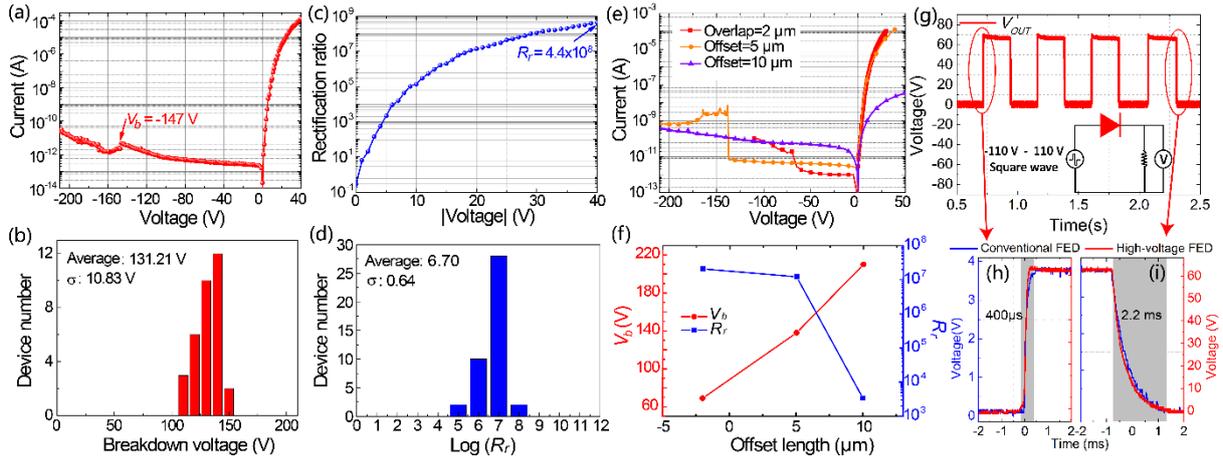

**Figure 3**. Electrical characteristics of flexible high-voltage diodes. (a) *I-V* characteristics with a high $V_b$ of 147 V. (b) Statistical data of $V_b$ with an average value of 131.21 V and standard deviation of 10.83 V. (c) $R_r$ versus voltage diagram with a high $R_r$ of 4.4× 10$^8$. (d) Statistical data of log ($R_r$) with an average value of 6.70 and standard deviation of 0.64. (e) *I-V* characteristics of diodes with 2 μm-overlap, 5 μm-offset and 10 μm-offset lengths, respectively. (f) Opposite tendency of $V_b$ and $R_r$ with regard to the offset length. (g) High-voltage square wave after rectification. Inset shows the circuit diagram. (h) Comparison of the rising times between conventional and high-voltage FEDs. (i) Comparison of the falling times between conventional and high-voltage FEDs.

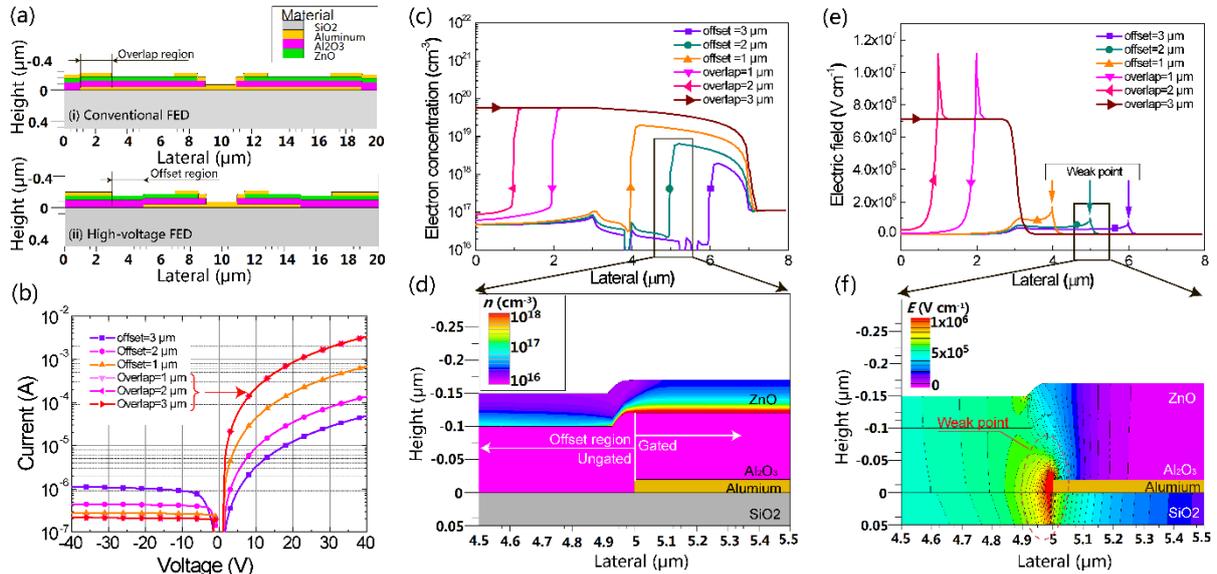

**Figure 4.** TCAD simulation of the high-voltage diode working principle. (a) Schematic of the overlap and offset structures. (b) Simulated *I-V* characteristics, in which diodes with longer offset region show less control of voltage over current. (c) Extracted electron concentration profiles along cutline 1 (height = -0.121 μm, lateral = 0-8 μm) under a positive bias. (d) Contour of electron concentration in 2 μm-offset diodes. The gated region is with high-electron concentration, while the ungated region with low-electron concentration. (e) Extracted electric field profiles along cutline 2 (height = -0.021 μm, lateral = 0-8 μm) under a negative bias. Electric field strength is largely mitigated by the offset region. (f) Contour of electric field in 2 μm-offset diodes. The weak point locates near the electrode edge.

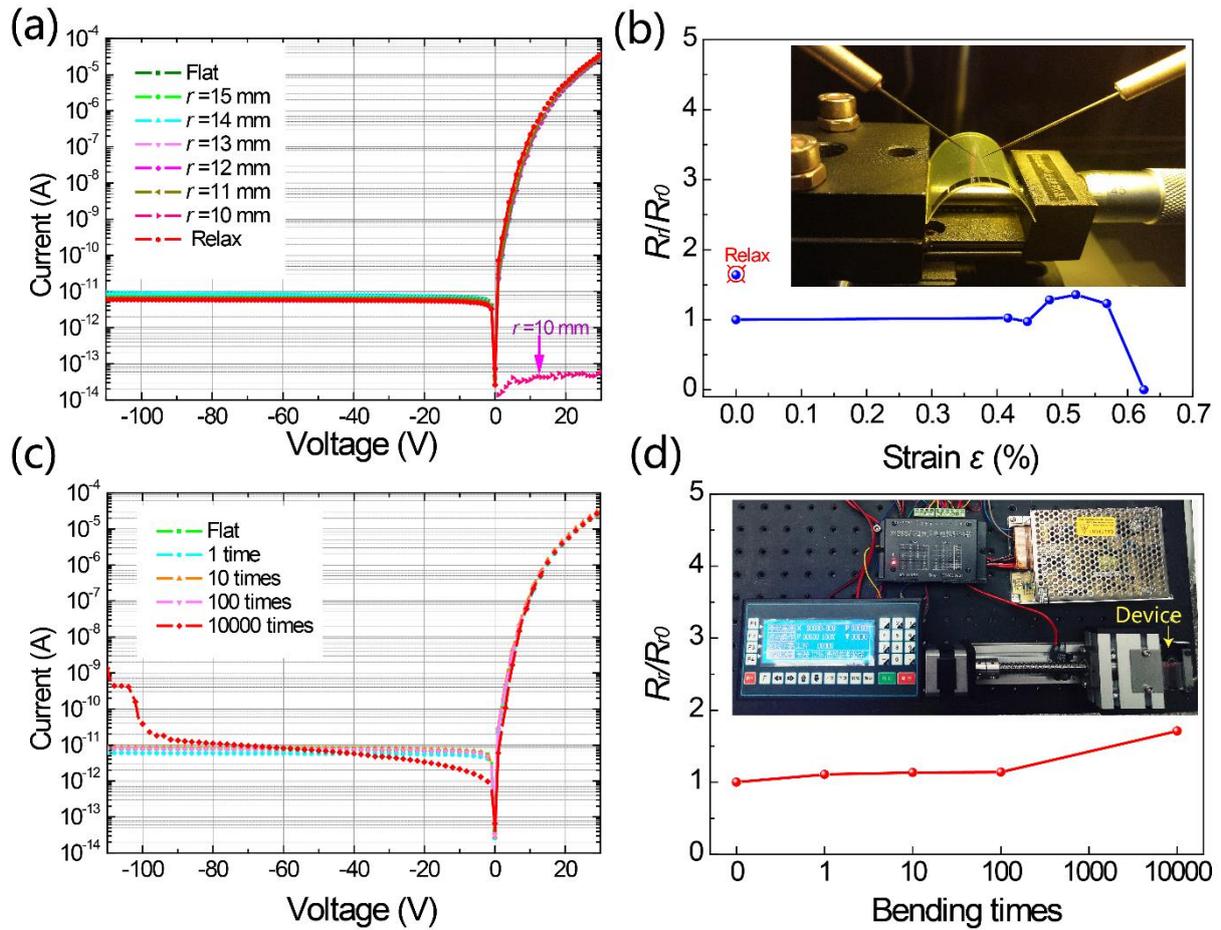

**Figure 5.** Mechanical bending tests of flexible high-voltage diodes. (a) *I-V* characteristics while substrate was flat and bent at various radii, respectively, which shows a safe operation limit of $\varepsilon = 0.57\%$ ($r = 11$ mm). (b) Variation of $R_r$ as a function of tensile stain. Inset shows the photograph of device under bending test. (c) Fatigue test of the high-voltage diode, in which little degradation of $V_b$ occurs after 10000 times bending. (d) Variation of $R_r$ as a function of bending times. Inset is the photograph of the self-assembled fatigue test system.

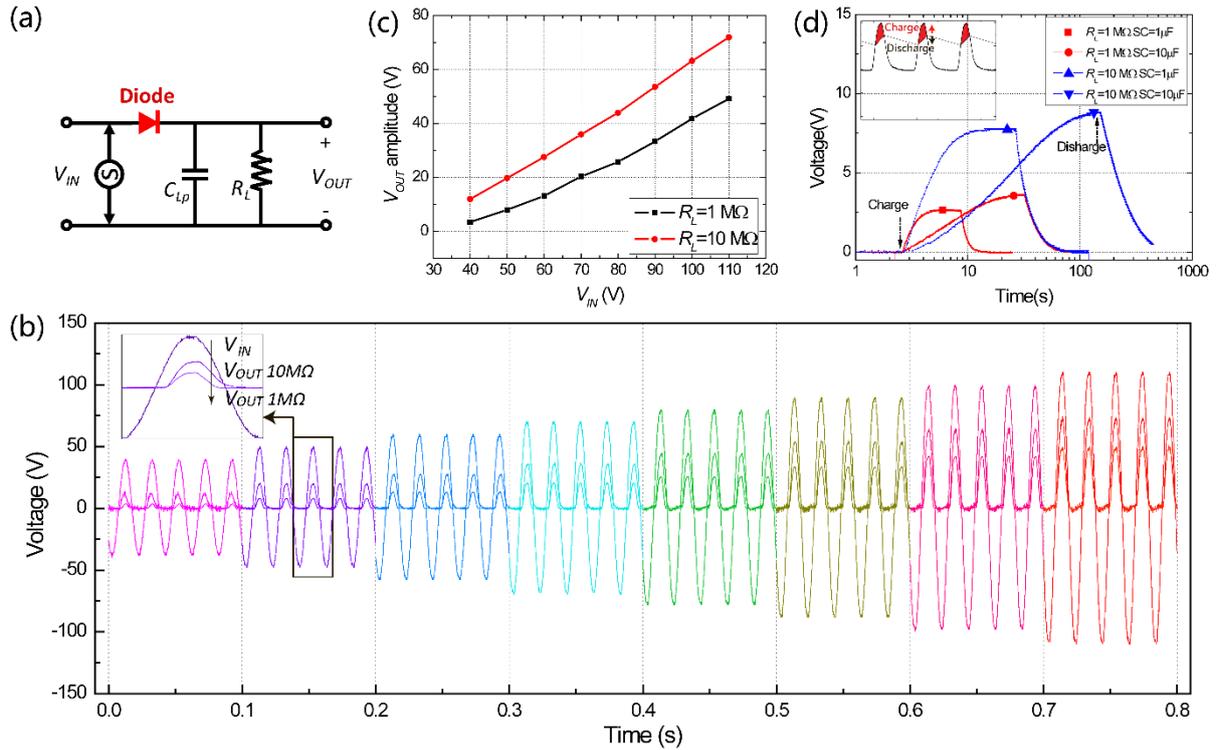

**Figure 6.** Half-wave rectifier circuit with one single flexible high-voltage diode. (a) Half-wave rectifier circuit diagram. (b) Waveforms of $V_{IN}$ and $V_{OUT}$ with 1 MΩ and 10 MΩ load resistors, respectively. (c) $V_{OUT}$ amplitude as a function of $V_{IN}$, which shows that $V_{OUT}$ is dependent both on $V_{IN}$ and $R_L$. (d) Charge and discharge of 1μF and 10 μF SCs with 1 MΩ and 10 MΩ load resistors. The charge and discharge behaviors follow RC circuit rules. The inset depicts that the charge behavior only occurs when $V_{OUT}$ is higher than the SC voltage, otherwise, the SC is discharged by the load resistor.

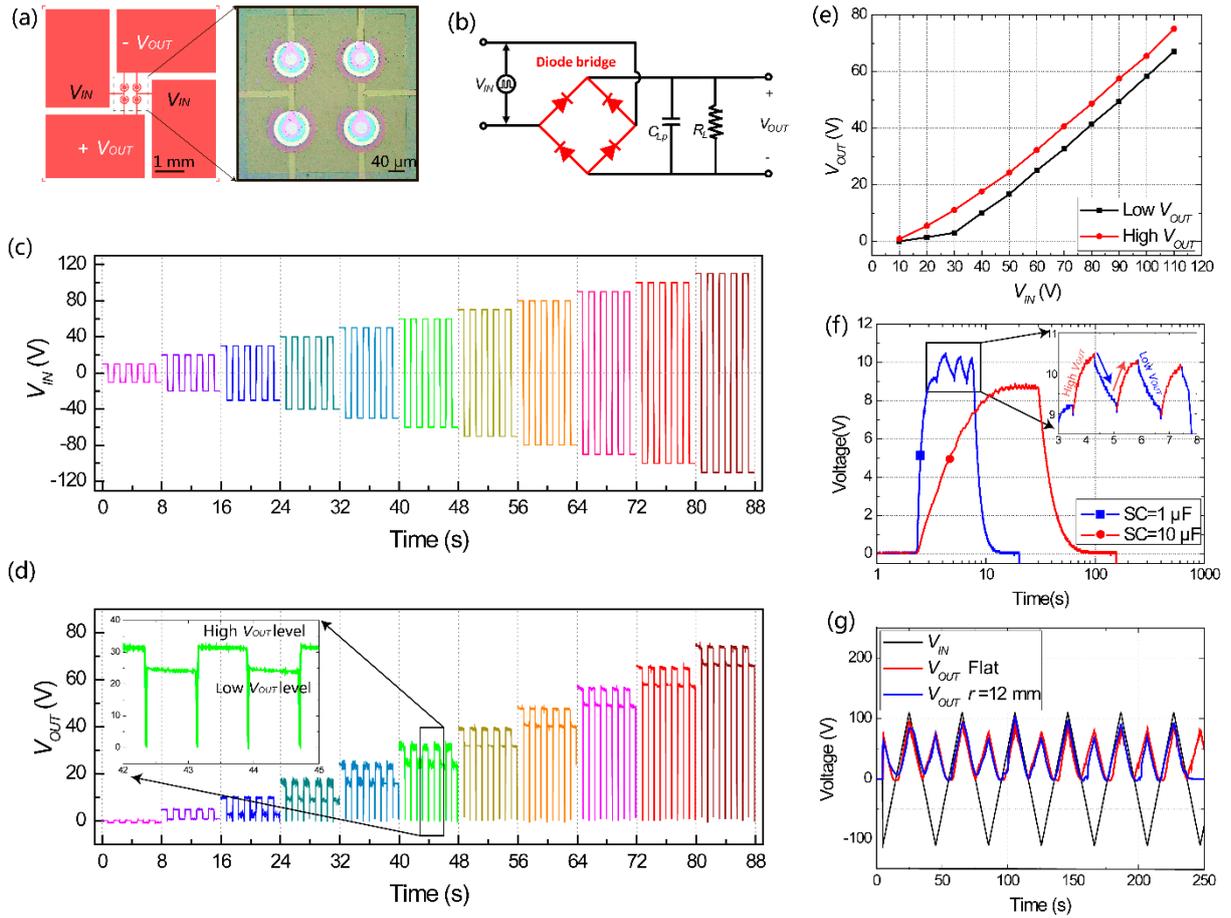

**Figure 7.** Full-wave rectifier circuit. (a) Device pattern and micro photograph of the four-diode rectifier bridge. (b) Full-wave rectifier circuit diagram. (c) $V_{IN}$ of the full-wave rectifier circuit. (d) $V_{OUT}$ of the full-wave rectifier circuit. Inset shows the high and low $V_{OUT}$ levels. (e) High and low $V_{OUT}$ levels as a function of $V_{IN}$. (f) Charge and discharge of 1 μF and 10 μF SCs. The inset shows a jagged saturation voltage of SC due to the charge and discharge behaviors with high and low $V_{OUT}$. (g) Full-wave rectification of triangle-wave high AC voltage while the substrate was flat and bent at $r$ = 12 mm, respectively.

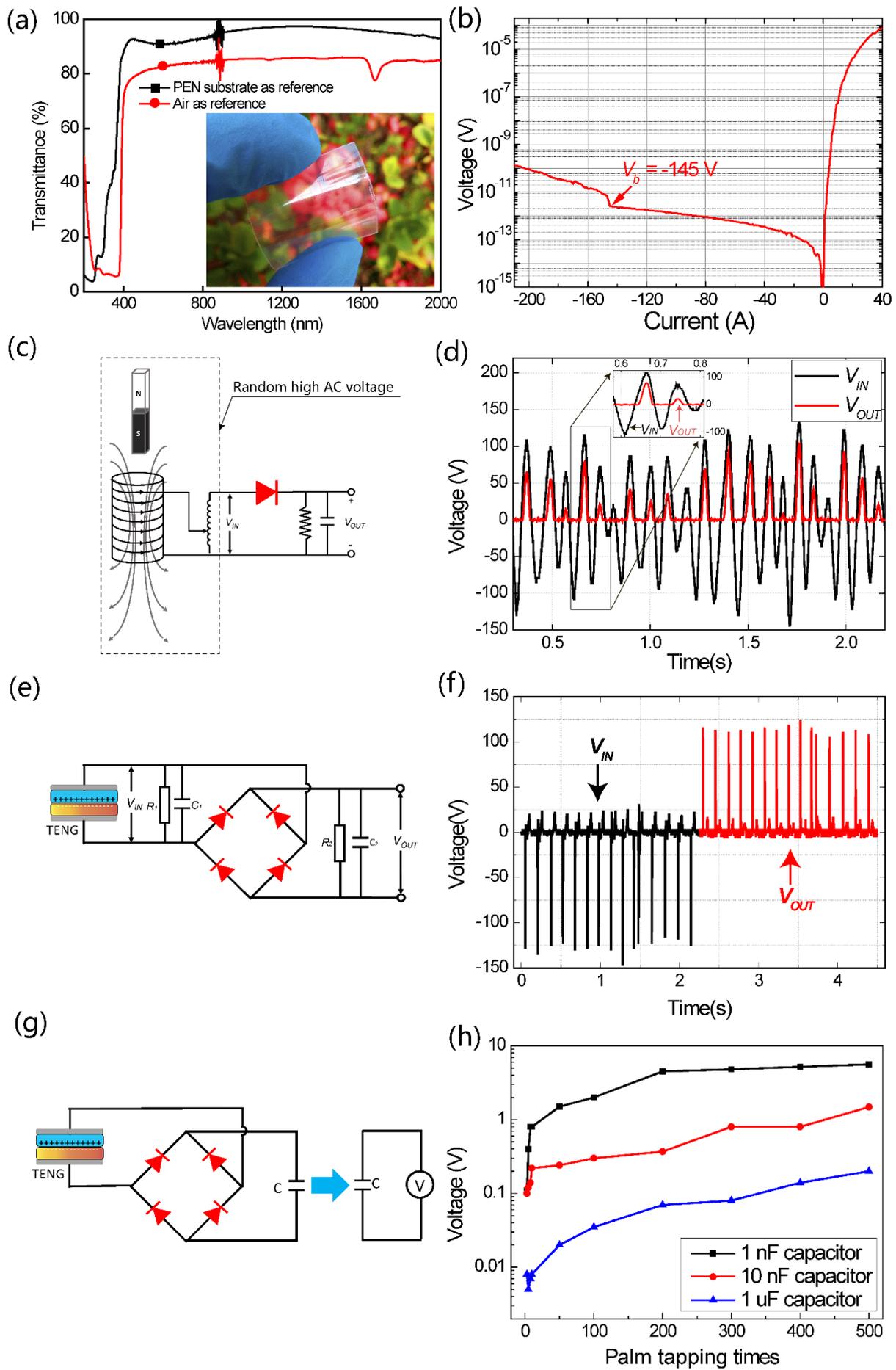

**Figure 8.** Flexible fully transparent high-voltage diodes. (a) Optical transmittance spectra with high transmittance over 80% in visible spectrum range. The inset is the photograph of the devices. (b) *I-V* characteristics with high $R_r$ ($2.18 \times 10^8$) and high $V_b$ (145 V). (c) Half-wave rectifier circuit diagram with EMG as power source. (d) $V_{IN}$ and $V_{OUT}$ of the rectifier circuits, which demonstrates the compatibility of present high-voltage diodes with EMGs. (e) Full-wave rectifier circuit diagram with TENG as power source. (f) $V_{IN}$ and $V_{OUT}$ of the rectifier circuits, which demonstrates the compatibility of present high-voltage diodes with TENGs. (g) Circuit diagram for charging capacitors with TENG and diode bridge. (h) Capacitor voltage versus palm tapping times, which demonstrate a complete energy supply strategy with energy harvest, management and storage.

**Vitae**

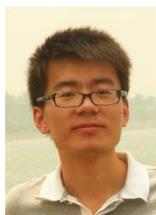

Yonghui Zhang received his Bachelor's degree in Physics from Shandong Normal University (SDNU) in 2012 and Master's degree in Material Engineering from Institute of Physics, Chinese Academy of Sciences (IOPCAS) in 2015. Now he is pursuing his Ph.D. in Condensed Matter Physics under the supervision of Prof. Zengxia Mei and Prof. Xiaolong Du in Renewable Energy Laboratory (REL), IOPCAS. His current research interest mainly focuses on flexible electronics.

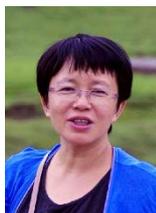

Zengxia Mei obtained her Ph.D. degree in IOPCAS under the supervision of Prof. Qikun Xue and Prof. Xiaolong Du in 2005. She joined the faculty of IOPCAS in 2005 and has been working in REL, IOPCAS from 2009 up to now. Her research interest is defects energetics in functional oxide semiconductors and devices as well as flexible optoelectronic and microelectronic devices.

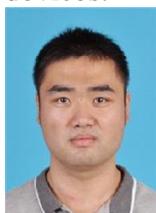

Tao Wang received his Bachelor's degree from SDNU in 2013 and Master's degree from Beijing Institute of Technology (BIT) in 2017. Now he is a research assistant in REL, IOPCAS. His current research interest focuses on TENGs.

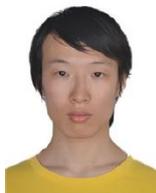

Wenxing Huo received his Bachelor's degree from Xidian University in 2012 and Master's degree from Beijing University of Posts and Telecommunications in 2015. Now he is pursuing his Ph.D. under the supervision of Prof. Zengxia Mei and Prof. Xiaolong Du in REL, IOPCAS. His current research interest focuses on flexible electronics.

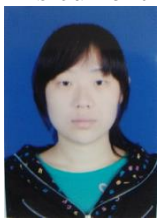

Shujuan Cui received her Bachelor's degree from Hebei University in 2013. Now she is pursuing her Ph.D. under the supervision of Prof. Zengxia Mei and Prof. Xiaolong Du in REL, IOPCAS. Her current research interest focuses on flexible photoelectronics.

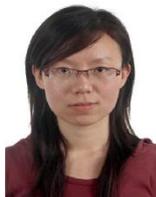

Huili Liang obtained her Ph.D. degree in IOPCAS under the supervision of Prof. Xiaolong Du and Prof. Zengxia Mei in 2012. She joined the faculty of IOPCAS in 2012 and has been working in REL, IOPCAS from 2012 up to now. Her current research interest focuses on flexible X-ray photodetectors.

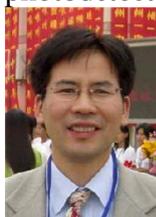

Xiaolong Du obtained his Ph.D. degree in BIT in 1999. He did post-doctoral research in Chiba University from 1999 to 2002. In 2002, he joined IOPCAS, where he is currently a Professor and group leader. His research interests include high-efficiency black silicon solar cells, ZnO based material science and device applications.

**Note:**
Color: online only.